\documentclass[floatfix,showpacs,aps,nofootinbib,twocolumn]{revtex4-1}
\usepackage{amssymb,amsmath}
\usepackage{graphicx,float}
\usepackage{indentfirst}
\usepackage{hyperref}

\newcommand{\beq}{\begin{equation}}
\newcommand{\eeq}{\end{equation}}
\newcommand{\bea}{\begin{eqnarray}}
\newcommand{\eea}{\end{eqnarray}}

\def\({\left(}
\def\){\right)}

\begin{document}

\title{The connection between Dirac dynamic and parity symmetry}
\author{C. H. Coronado Villalobos}
\email{ccoronado@feg.unesp.br}
\author{R. J. Bueno Rogerio}
\email{rodolforogerio@feg.unesp.br}
\affiliation{Departamento de F\'{\i}sica e Qu\'{\i}mica, Universidade Estadual Paulista,
Guaratinguet\'{a}, SP, Brazil.}


\begin{abstract}
Dirac spinors are important objects in the current literature, the algebraic structure presented in the text-books is a general method to write it, however, not unique. The purpose of the present work is to show an alternative approach to construct Dirac spinors, considering the interchange between the Lorentz representation space $(1/2,0)$ and $(0,1/2)$ made by the ``\textit{Magic of Pauli matrices}'' and not by parity, as commonly it was thought. As it is well known, parity operator is related with the Dirac dynamics, as it can be seen in the reference \cite{speranca}. The major focus is to establish the relation between Dirac dynamics with parity operator, the reverse path shown in \cite{speranca}.  
\end{abstract}


\maketitle

\section{Introduction}
It is known that the full Lorentz group is composed by rotation generators, $\boldsymbol{J}$, boost generators, $\boldsymbol{K}$, and discrete symmetries: parity, $P$, and time reversion, or time reflection, $T$ \cite{weinberg1}. 
The Dirac spinors are built within the full Lorentz group, keeping as the transformation law between the spinorial components, left-hand and right-hand components, the parity symmetry. In other words, this symmetry is exactly the link element between both parts of the representation space, i.e., it connects the subspaces $(j,0)$ and $(0,j)$. In reference \cite{speranca} was shown that parity operator, $P$, in the spin $1/2$ representation is given by
\begin{eqnarray}
P\psi(\boldsymbol{p}) = m^{-1}\gamma_{\mu}p^{\mu}\psi(\boldsymbol{p}),
\end{eqnarray} 
which allows us to derive the Dirac equation from the space-time symmetries alone.

Nevertheless, there are cases where the representation spaces are not connected by discrete symmetries, like in the $Elko´s$ spinors case, proposed in reference \cite{jcap}.
In this case the transformation law between the spinorial components is known as ``\textit{Magic of Pauli matrices}''; this observation was first made by Ramond in \cite{ramond}, and it reads

\begin{eqnarray}\label{paulimatrices}
\Theta \sigma \Theta^{-1} = -\sigma^{*},
\end{eqnarray}
where $\Theta$, in the spin $1/2$ representation, is given by
\begin{eqnarray}
\Theta = \left(\begin{array}{cc}
0 & -1 \\ 
1 & 0
\end{array}\right).
\end{eqnarray}
The above operator is known by \emph{Wigner Time Reversal Operator}.

Following the Elko recipe, we present a set of single helicity spinors, which endows the ``\textit{Magic of Pauli matrices}'' as a crucial link between the representation space, for this case $(1/2,0)$ and $(0,1/2)$. The relevance of this work lies in the consequence of once imposed the Dirac dynamics for the spinors, it takes us directly to a relation between the phases and concomitantly the same relation is observed when it is imposed to the spinors being eigenspinors of parity operator. On the other hand, if the spinors satisfies the Dirac dynamics accordingly they are eigenspinors of parity operator, and as was shown in \cite{speranca} the inverse path is truth too.

This paper is organized as it follows: In Section II we define dual helicity spinor as well single helicity spinors and we present the right and left hand components and its respective helicity. In Section III we analyse the behaviour of the spinors under action of the Dirac operator, via two distinct methods, and then the phases fixation to be coherent with the Dirac dynamics, finally we show that also it satisfies the Klein-Gordon equation. Section IV is reserved to ascertain the action of parity symmetry over the spinors. We construct the dual spinor in Section V aiming to write the spin sums.

\section{Proposing new spinors}
Motivated by the Elko theoretical discovery \cite{prd72}, we present here the construction of a spinor which inherits the same law of transformation between the spinorial components, or in other words, the ``\textit{magic of the Pauli matrices}''.
Following the same way of thought used in the Elko construction, we are able to obtain four spinors; two of them are dual helicity objects and the remaining two are single helicity objects. So, we start by showing the dual helicity spinors. Basically these spinors essentially must be recognized as Elko. They are defined by  
\begin{eqnarray}
\label{3}\psi^{\lbrace +,+ \rbrace}(\boldsymbol{0})\equiv \left(\begin{array}{c}
\varsigma_1 \Theta\phi_L^{*\lbrace + \rbrace}(\boldsymbol{0}) \\
\varsigma_2 \Theta\phi_R^{*\lbrace + \rbrace}(\boldsymbol{0})
\end{array} \right), 
\\
\label{331}\psi^{\lbrace -,- \rbrace}(\boldsymbol{0})\equiv \left(\begin{array}{c}
\varsigma^{\prime}_1 \Theta\phi_L^{*\lbrace - \rbrace}(\boldsymbol{0}) \\ 
\varsigma^{\prime}_2 \Theta\phi_R^{*\lbrace - \rbrace}(\boldsymbol{0})
\end{array}\right),
\end{eqnarray}
where $\varsigma\in\mathbb{C}$ are arbitrary phases. For Elko spinors, the phases are fixed imposing $C\psi^{h}=\pm\psi^{h}$ \cite{jcap}. For the single helicity spinors, we have the following representation
\begin{eqnarray}
\label{4}\psi^{\lbrace +,- \rbrace}(\boldsymbol{0})\equiv \left(\begin{array}{c}
\xi_1 \Theta\phi_L^{*\lbrace + \rbrace}(\boldsymbol{0}) \\ 
\xi_2 \Theta\phi_R^{*\lbrace - \rbrace}(\boldsymbol{0})
\end{array}  \right),\\ 
\label{44444}\psi^{\lbrace -,+ \rbrace}(\boldsymbol{0})\equiv \left(\begin{array}{c}
\xi^{\prime}_1 \Theta\phi_L^{*\lbrace - \rbrace}(\boldsymbol{0}) \\ 
\xi^{\prime}_2 \Theta\phi_R^{*\lbrace + \rbrace}(\boldsymbol{0})
\end{array}  \right),
\end{eqnarray}
the phases $\xi\in\mathbb{C}$ represents an arbitrary phase. Using the ``\textit{magic of the Pauli matrices}'' it allows to write, with no mention to discrete symmetries, a complete set of dual and single helicity spinors. In the present work, our aim is to use this mathematical device to study the possibility to construct Dirac spinors, changing the usual link between the representation space $(1/2,0)$ and $(0,1/2)$, usually made via parity, but now made by relation \eqref{paulimatrices}. In this fashion, looking forward to understand the physical content in this new set of spinors and study any consequence by the absent of discrete symmetry. 

Choosing the components to be eigenstates of the helicity operator, $\vec{\sigma}\cdot\hat{p}$, one finds the left-hand spinorial componentes at the rest frame \cite{jcap, helicidade} 
\begin{eqnarray}\label{compesq}
\phi_L^{\lbrace+\rbrace}(\boldsymbol{0}) = \sqrt{m}\left(\begin{array}{c}
\cos(\theta/2)e^{-i\phi/2} \\ 
\sin(\theta/2)e^{i\phi/2}
\end{array} \right), \\ \phi_L^{\lbrace-\rbrace}(\boldsymbol{0}) = \sqrt{m}\left(\begin{array}{c}
-\sin(\theta/2)e^{-i\phi/2} \\ 
\cos(\theta/2)e^{i\phi/2}
\end{array} \right),
\end{eqnarray}
and the right-hand components as well
\begin{eqnarray}\label{compdir}
\phi_R^{\lbrace+\rbrace}(\boldsymbol{0}) = \sqrt{m}\left(\begin{array}{c}
-i\sin(\theta/2)e^{-i\phi/2} \\ 
i\cos(\theta/2)e^{i\phi/2},
\end{array} \right), \\ \phi_R^{\lbrace-\rbrace}(\boldsymbol{0}) = \sqrt{m}\left(\begin{array}{c}
-i\cos(\theta/2)e^{-i\phi/2} \\ 
-i\sin(\theta/2)e^{i\phi/2}
\end{array} \right).
\end{eqnarray}
The mass term is chosen such that in the massless limit the rest spinors in the representation space $(1/2,0)$ and $(0,1/2)$ identically vanish, so that there can be no massless particle at rest \cite{majoranalike}.
The interaction amplitudes must have the factor $m^{j}$, where $j$ is the spin of the particle, in order to the $m \rightarrow 0$ limit be consistent \cite{marinov}.

Regarding the helicity, we have to described the action of the operator $\vec{\sigma}\cdot\hat{p}$ at the spinorial components. Thus, we can define the following relations
\begin{eqnarray}\label{relacaohelicidade}
\vec{\sigma}\cdot\hat{p}\,\,\Theta\phi_L^{*\lbrace\pm\rbrace}(\boldsymbol{0}) = \mp\Theta\phi_L^{*\lbrace\pm\rbrace}(\boldsymbol{0}), \\
\vec{\sigma}\cdot\hat{p}\,\,\Theta\phi_R^{*\lbrace\pm\rbrace}(\boldsymbol{0}) = \pm\Theta\phi_R^{*\lbrace\pm\rbrace}(\boldsymbol{0}).
\end{eqnarray}
Such relationships are of great importance and will be used in the scope of this work.
Once we had defined above the action of the operator $\vec{\sigma}\cdot\hat{p}$ on the rest-frame spinors, one is able to write the spinors in a boosted-frame with an arbritary momentum. So, the boosted spinors are given by the action of the boost matrix, as we can see below
\begin{eqnarray}\label{matrizboost}
\psi^{\lbrace h,-h\rbrace}(\boldsymbol{p})= \left(\begin{array}{cc}
\kappa^{(1/2,0)} & 0 \\ 
0 & \kappa^{(0,1/2)}
\end{array}\right) \psi^{\lbrace h,-h\rbrace}(\boldsymbol{0}),
\end{eqnarray}
where the boost matrix components $\kappa^{(1/2,0)}$ and $\kappa^{(0,1/2)}$ are the right-hand and left-hand boost operators, respectively. They are defined by \cite{jcap,plb687}
\begin{eqnarray}
\kappa^{(1/2,0)} &=& \sqrt{\frac{E+m}{2m}}\big(\mathbb{I} + \frac{\vec{\sigma}.\vec{p}}{E+m}\big),\\
\kappa^{(0,1/2)} &=& \sqrt{\frac{E+m}{2m}}\big(\mathbb{I} - \frac{\vec{\sigma}.\vec{p}}{E+m}\big).
\end{eqnarray}
In this way, the boosted-frame dual helicity spinors are given by
\begin{eqnarray}
\label{boostdualmais}\psi^{\lbrace +,+\rbrace}(\boldsymbol{p})&=&\sqrt{\frac{E+m}{2m}}\big(1 - \frac{p}{E+m}\big)\psi^{\lbrace +,+\rbrace}(\boldsymbol{0}), \\
\label{boostdualmenos}\psi^{\lbrace -,-\rbrace}(\boldsymbol{p})&=&\sqrt{\frac{E+m}{2m}}\big(1 + \frac{p}{E+m}\big)\psi^{\lbrace -,-\rbrace}(\boldsymbol{0}),
\end{eqnarray} 
and for the single helicity case we have
\begin{eqnarray}\label{19}
\psi^{\lbrace h \rbrace}(\boldsymbol{p}) = \sqrt{\frac{E+m}{2m}}\left(\begin{array}{cc}
\mathbb{I}+ p\frac{\vec{\sigma}.\hat{p}}{E+m} & 0 \\ 
0 & \mathbb{I}- p\frac{\vec{\sigma}.\hat{p}}{E+m}
\end{array} \right)\psi^{\lbrace h \rbrace}(\boldsymbol{0}), \!\!\!\!\!\!\!\!\!\!\!\!\!\!\!\!\nonumber\\ 
\end{eqnarray}
where $h$ stands for ${\lbrace \pm,\mp\rbrace}$ helicities. So, in this manner, it is possible to define all the possible spinors in an arbitrary momentum inertial frame. The Dirac equation can be derived by simply appealing to the properties of the parity operator $P$. So we define such operator to be \begin{eqnarray}\label{pari}
P\psi(\boldsymbol{p}) = \eta\psi(\boldsymbol{-p}),
\end{eqnarray}
where $\eta$ an off-diagonal matrix
\begin{eqnarray}
\eta = \left(\begin{array}{cc}
0 & \mathbb{I} \\ 
\mathbb{I} & 0
\end{array}  \right).
\end{eqnarray} 
Then, using \eqref{19} and \eqref{pari}, we easily obtain 
\begin{eqnarray}
P = m^{-1}\gamma_{\mu}p^{\mu}.
\end{eqnarray}
The above operator has eigenvalues $\pm 1$ \cite{speranca}. Indeed, the focus of this paper, is to find spinors governed by the Dirac dynamic and then verify its behaviour under parity operator, showing the inverse path and the correspondence between the mentioned method and the method analysed by L. Speran\c{c}a in reference \cite{speranca}.

\section{Field dynamics}
In this section, our focus is to verify the dynamic associated to the proposed spinors. The set of spinors belongs to Lorentz proper orthochronous group, formed only by rotation and boost generators; by this reason the focus is now to evaluate any emergent consequences, if any carried by the mentioned fact, in the dynamical interpretation. For this task we start analysing whether this set of spinors satisfy Dirac equation and how the phases must be related and fixed. Finally, we shall evaluate if it satisfies Klein-Gordon equation, as a consistent check.

In order to verify if all the spinors satisfies Dirac equation, we employ here two distinct methods. Start using the same mathematical construction that was made in \cite{ryder} and lastly the second method is the action of the Dirac operator, $\gamma_{\mu}p^{\mu}$, upon the spinors. 

Here we start using the same formalism used in \cite{ryder}, where the Dirac operator is built starting from a rest-frame spinor and then acting with the boost operator, from \eqref{matrizboost} it is known that  
\begin{eqnarray}\label{995}
\!\!\!\psi^{\lbrace +,-\rbrace}(\boldsymbol{p}) = \!\!\sqrt{\frac{E+m}{2m}}\!\!\left(\begin{array}{cc}
\mathbb{I}+ p\frac{\vec{\sigma}.\hat{p}}{E+m} & 0 \\ 
0 & \mathbb{I}- p\frac{\vec{\sigma}.\hat{p}}{E+m}
\end{array} \right)\!\!\psi^{\lbrace +,-\rbrace}(\boldsymbol{0}).\!\!\!\!\!\!\!\!\!\!\!\!\!\!\!\!\!\!\!\!\!\!\!\nonumber\\
\end{eqnarray}
For didactic purposes, as an example, we use $\psi^{\lbrace +,-\rbrace}(\boldsymbol{p})$  \footnote{In the present approach we use only $\psi^{\lbrace +,-\rbrace}(\boldsymbol{p})$, to simplify the calculations. The algorithm is identical to the other spinors.}
\begin{eqnarray}
\!\!\!\!\!\!\psi^{\lbrace +,-\rbrace}(\boldsymbol{p}) = \sqrt{\frac{E+m}{2m}}\left(\!\!\! \begin{array}{c}
\big(\mathbb{I}+ p\frac{\vec{\sigma}.\hat{p}}{E+m}\big)\xi_1 \Theta\phi_L^{*\lbrace +\rbrace}(\boldsymbol{0}) \\ 
\big(\mathbb{I}- p\frac{\vec{\sigma}.\hat{p}}{E+m}\big)\xi_2 \Theta\phi_R^{*\lbrace -\rbrace}(\boldsymbol{0})
\end{array}\!\!\! \right).\!\!\!\!\!\!\!\!\!\!\!\!\!\!\!\!\!\!\!\!\!\!\!\!\!\!\!\!\!\!\!\!\nonumber\\
\end{eqnarray}
where we can easily identify the right-hand component as  
\begin{eqnarray}\label{opdireita}
\phi_R^{\lbrace +\rbrace}(\boldsymbol{p}) = \frac{E+m+\vec{\sigma}.\vec{p}}{\sqrt{2m(E+m)}}\xi_1\Theta\phi_L^{*\lbrace +\rbrace}(\boldsymbol{0}),
\end{eqnarray}
with the left-hand component is
\begin{eqnarray}\label{opesquerda}
\phi_L^{\lbrace -\rbrace}(\boldsymbol{p}) = \frac{E+m-\vec{\sigma}.\vec{p}}{\sqrt{2m(E+m)}}\xi_2\Theta\phi_R^{*\lbrace -\rbrace}(\boldsymbol{0}).
\end{eqnarray}
As the next step, we present some usefull relations:   
\begin{eqnarray}
\label{41}\Theta\phi_L^{*\lbrace + \rbrace}(\boldsymbol{0}) &=& -i\Theta\phi_R^{*\lbrace - \rbrace}(\boldsymbol{0}), \\
\label{42}\Theta\phi_R^{*\lbrace - \rbrace}(\boldsymbol{0}) &=& i\Theta\phi_L^{*\lbrace + \rbrace}(\boldsymbol{0}).
\end{eqnarray}
Inserting the last two relations above in the equations \eqref{opdireita} and \eqref{opesquerda}, it is easy to find  
\begin{eqnarray}\label{666}
\phi_R^{\lbrace +\rbrace}(\boldsymbol{p}) = -i\xi_1\xi_2^*\big[\frac{E+\vec{\sigma}.\vec{p}}{m}\big]\phi_L^{\lbrace -\rbrace}(\boldsymbol{p}),
\end{eqnarray}
and 
\begin{eqnarray}\label{667}
\phi_L^{\lbrace -\rbrace}(\boldsymbol{p}) = i\xi_1^*\xi_2\big[\frac{E-\vec{\sigma}.\vec{p}}{m}\big]\phi_R^{\lbrace +\rbrace}(\boldsymbol{p}).
\end{eqnarray}
The formalism presented in these calculations shows that equations \eqref{666} and \eqref{667} allows to write them in a simple matricial form 
\begin{eqnarray}\label{diracespinor}
\!\!\!\!\!\!\!\!\!\left(\begin{array}{cc}
m & -i\xi_1\xi^{*}_2(p_0+\vec{\sigma}.\vec{p}) \\ 
i\xi^{*}_1\xi_2(p_0-\vec{\sigma}.\vec{p}) & -m
\end{array}  \right)\!\!\!\left(\!\!\! \begin{array}{c}
\phi_R^{\lbrace +\rbrace}(\boldsymbol{p}) \\ 
\phi_L^{\lbrace -\rbrace}(\boldsymbol{p})
\end{array}\!\!\!\right) = 0.\!\!\!\!\!\!\!\!\!\!\!\!\!\!\!\!\!\!\!\!\!\!\!\!\!\!\!\!\!\!\!\!\nonumber\\
\end{eqnarray}
In order to the last equation be regarded as Dirac equation, we must fix the phases as
\begin{eqnarray}
\xi_1 = i\xi_2.
\end{eqnarray} 
With this condition, equation \eqref{diracespinor} becomes 
\begin{eqnarray}\label{diracespinor2}
\left(\begin{array}{cc}
-m & (p_0+\vec{\sigma}.\vec{p}) \\ 
(p_0-\vec{\sigma}.\vec{p}) & -m
\end{array}  \right)\left(\begin{array}{c}
\phi_R^{\lbrace +\rbrace}(\boldsymbol{p}) \\ 
\phi_L^{\lbrace -\rbrace}(\boldsymbol{p})
\end{array}  \right) = 0,
\end{eqnarray}
making possible to write \eqref{diracespinor2} as 
\begin{eqnarray}\label{1072}
(\gamma_{\mu}p^{\mu}-m\mathbb{I})\psi^{\lbrace +, - \rbrace}(\boldsymbol{p}) = 0.
\end{eqnarray}
Repeating the same procedure for $\psi^{\lbrace -,+ \rbrace}(\boldsymbol{p})$, we conclude 
\begin{eqnarray}\label{26666}
\xi^{\prime}_1 = -i\xi^{\prime}_2,
\end{eqnarray}
in this such manner
\begin{eqnarray}\label{108}
(\gamma_{\mu}p^{\mu}-m\mathbb{I})\psi^{\lbrace -, + \rbrace}(\boldsymbol{p}) = 0.
\end{eqnarray}
The last equation evidences to the reader that all $\psi(\boldsymbol{p})$ satisfies Dirac equation, as far as Eq. \eqref{26666} holds . Applying once again the same procedure in the equation \eqref{108}, it becomes easy to verify 
\begin{eqnarray}
(\Box + m^2)\psi^{\lbrace \pm,\mp\rbrace}(\boldsymbol{p}) = 0.
\end{eqnarray}
Which reproduces Klein-Gordon equation annihilating all $\psi^h(\boldsymbol{p})$.

As said before, the second method consists in analyse spinors under action of the $\gamma_{\mu}p^{\mu}$ operator, taking care with conditions that governs the phases terms. In this way, we operate with $\gamma_{\mu}p^{\mu}$ upon the spinors in  \eqref{4}, thus
\begin{eqnarray}\label{102}
\gamma_{\mu}p^{\mu}\psi^{\lbrace +,-\rbrace}(\boldsymbol{p}) = \bigg[E\gamma_0 + \left(\begin{array}{cc}
0 & p\vec{\sigma}.\hat{p} \\ 
-p\vec{\sigma}.\hat{p} & 0
\end{array} \right) \bigg] \psi^{\lbrace +,-\rbrace}(\boldsymbol{p}).\!\!\!\!\!\!\!\!\!\!\!\!\!\!\!\!\!\!\!\!\!\!\!\!\!\!\!\!\!\!\!\!\nonumber\\
\end{eqnarray}
Using equations \eqref{relacaohelicidade} and \eqref{995}, one is able to write the spinor in a boosted-frame with an arbitrary momentum as \footnote{In order to summarize the notation we define the boosts operators as it reads: $\Lambda_{\pm}(p^{\mu}) \equiv \sqrt{\frac{E+m}{2}}\bigg(1 \pm p\frac{\boldsymbol{\sigma}\cdot\hat{p}}{E+m}\bigg)$.}
\begin{eqnarray}\label{104}
\psi^{\lbrace +,-\rbrace}(\boldsymbol{p}) = \left(\begin{array}{c}
\xi_1\Lambda_{-}(p^{\mu})\Theta\phi_L^{*\lbrace +\rbrace}(\boldsymbol{0}) \\ 
 \\ 
\xi_2\Lambda_{+}(p^{\mu})\Theta\phi_R^{*\lbrace -\rbrace}(\boldsymbol{0})
\end{array}\right). 
\end{eqnarray}
Now, operating with $\gamma_{\mu}p^{\mu}$ over \eqref{104} we have 
\begin{eqnarray}\label{105}
\gamma_{\mu}p^{\mu}\psi^{\lbrace +,-\rbrace}(\boldsymbol{p}) =  \left(\begin{array}{c}
(E-p)\xi_2\Lambda_{+}(p^{\mu})\Theta\phi_R^{*\lbrace - \rbrace}(\boldsymbol{0}) \\ 
• \\ 
(E+p)\xi_1\Lambda_{-}(p^{\mu})\Theta\phi_L^{*\lbrace + \rbrace}(\boldsymbol{0})
\end{array}  \right)\!\!,
\end{eqnarray}
and we can see that $\psi^{\lbrace +,-\rbrace}(\boldsymbol{p})$ is not an eigenspinor of $\gamma_{\mu}p^{\mu}$. It is easy to see that the matricial components of equation \eqref{105} is shifted when compared to \eqref{104}.  
To continue with the calculations, we use the following trick, actually, we are using the relations between $\Theta$ and the spinorial left and right-hand components, given in equations \eqref{41} and \eqref{42}.

Taking that into account and using the last two relations above, one is able to write
\begin{eqnarray}\label{112}
\gamma_{\mu}p^{\mu}\psi^{\lbrace +,-\rbrace}(\boldsymbol{p}) = im\left( \begin{array}{c}
\xi_2\Lambda_{-}(p^{\mu})\Theta\phi_L^{*\lbrace + \rbrace}(\boldsymbol{0}) \\ 
• \\ 
-\xi_1\Lambda_{+}(p^{\mu})\Theta\phi_R^{*\lbrace - \rbrace}(\boldsymbol{0}),
\end{array} \right).
\end{eqnarray}
Firstly, for $\psi^{\lbrace +,-\rbrace}(\boldsymbol{p})$, equation \eqref{112} only represents Dirac equation, if the following conditions are imposed over the phases 
\begin{eqnarray}\label{dirac+-}
\xi_1 = + i\xi_2 \quad\mbox{and} \quad
\xi_2 = - i\xi_1,
\end{eqnarray}
and for $\psi^{\lbrace -,+\rbrace}(\boldsymbol{p})$, relation must be setted as
\begin{eqnarray}\label{dirac-+}
\xi^{\prime}_1 = - i\xi^{\prime}_2 \quad\mbox{and} \quad
\xi^{\prime}_2 = + i\xi^{\prime}_1,
\end{eqnarray} 
which is in agreement with first method, thus, showing the consistency between both methods. As it was expected, the spinors must to satisfy Klein-Gordon equation, then we easily conclude
\begin{eqnarray}
(\Box + m^2)\psi^{\{+,-\}}(\boldsymbol{p}) &=& 0.
\end{eqnarray}
Obviously the last equation is satisfied by the other remaining spinors. Thereby, we have showed that both methods works and are consistent.

\section{Defining a complete set of spinors via parity operator}

In order to built a complete set for the $\psi$ spinors, we introduce here the action of parity operator, and like as Dirac case. In the $(1/2,0)\oplus (0,1/2)$ representation space, the parity operator reads
\begin{equation}
P = e^{i\Phi}\gamma_0 \mathcal{R},
\end{equation}
where the $\mathcal{R}$ acts as
\begin{eqnarray}
\mathcal{R} \equiv \lbrace \theta \rightarrow \pi -\theta, \phi \rightarrow \phi+\pi, p \rightarrow p \rbrace ,
\end{eqnarray}
and we consider an arbitraty phase $e^{i\Phi}$, which will be fixed if necessary. Thus, we act with the parity operator over $\psi^{\lbrace -,+\rbrace}(\boldsymbol{p})$  
 \begin{eqnarray}\label{127}
P\psi^{\lbrace -,+\rbrace}(\boldsymbol{p}) = e^{i\Phi}\gamma_0 \mathcal{R} \left( \begin{array}{c}
\xi^{\prime}_1\Lambda_{+}(p^{\mu})\Theta\phi_L^{*\lbrace - \rbrace}(\boldsymbol{0}) \\ 
\xi^{\prime}_2\Lambda_{-}(p^{\mu})\Theta\phi_R^{*\lbrace + \rbrace}(\boldsymbol{0})
\end{array} \right).
\end{eqnarray}
Analysing the action of $\mathcal{R}$ over the spinors components, we have
\begin{eqnarray}
\mathcal{R}\phi_L^{*\lbrace - \rbrace}(\boldsymbol{p}) = \phi_L^{*\lbrace - \rbrace}(\boldsymbol{p}), \\
\mathcal{R}\phi_R^{*\lbrace + \rbrace}(\boldsymbol{p}) = \phi_R^{*\lbrace + \rbrace}(\boldsymbol{p}).
\end{eqnarray} 
Finally, the equation \eqref{127} can be expresed as 
\begin{eqnarray}\label{110}
P\psi^{\lbrace -,+\rbrace}(\boldsymbol{p}) = e^{i\Phi}\left(\begin{array}{c}
-i\xi^{\prime}_2\Lambda_{+}(p^{\mu})\Theta\phi_L^{*\lbrace -\rbrace}(\boldsymbol{0}) \\ 
i\xi^{\prime}_1\Lambda_{-}(p^{\mu})\Theta\phi_R^{*\lbrace +\rbrace}(\boldsymbol{0})
\end{array}  \right).
\end{eqnarray}
In order that the spinor on equation \eqref{110} becomes an eigenspinor of parity operator, we impose 
\begin{eqnarray}\label{relacaoparidade1}
\xi^{\prime}_2 &=& i\xi^{\prime}_1e^{i\Phi},
\end{eqnarray}
and for the other case, $\psi^{\lbrace +,-\rbrace}(\boldsymbol{p})$, reads
\begin{eqnarray}
\xi_2 &=& -i\xi_1e^{i\Phi},
\end{eqnarray}
fixing $e^{i\Phi}=+1$ we have positive eigenvalues, so it is easy then to conclude $P\psi^{\{\lambda\}}_{+}=+\psi^{\{\lambda\}}_{+}$. Now, imposing $\psi$ to have negative eigenvalues, we must to fix $e^{i\Phi}=-1$, in other words, $P\psi^{\{\lambda\}}_{-}=-\psi^{\{\lambda\}}_{-}$, then, for the phases of $\psi^{\lbrace +,-\rbrace}(\boldsymbol{p})$ we have the condition \footnote{To avoid confusion and to elucidate that the phases are different from the previous analysis, we made the following $\xi_1 \rightarrow \xi_3$ and $\xi_2 \rightarrow \xi_4$, the same still valid for the primed phases.},
\begin{eqnarray}\label{25}
\xi_4= i\xi_3,
\end{eqnarray}
and for the other case, $\psi^{\lbrace -,+\rbrace}(\boldsymbol{p})$, it reads
\begin{eqnarray}\label{26}
\xi^{\prime}_{4} = -i\xi^{\prime}_3,
\end{eqnarray}
So, we are able to write a complete set of single helicity spinors:
\begin{eqnarray}
\label{27}\psi^{\{+,-\}}_{+}(\boldsymbol{p}) &=& \left( \begin{array}{c}
\xi_1\Lambda_{-}(p^{\mu})\Theta\phi_{L}^{\{+\}*}(\boldsymbol{0}) \\ 
-i\xi_1\Lambda_{+}(p^{\mu})\Theta\phi_{R}^{\{-\}*}(\boldsymbol{0})
\end{array}  \right),\\\nonumber\\
\label{28}\psi^{\{-,+\}}_{+}(\boldsymbol{p}) &=& \left( \begin{array}{c}
\xi^{\prime}_1\Lambda_{+}(p^{\mu})\Theta\phi_{L}^{\{-\}*}(\boldsymbol{0}) \\ 
i\xi^{\prime}_1\Lambda_{-}(p^{\mu})\Theta\phi_{R}^{\{+\}*}(\boldsymbol{0})
\end{array}  \right),
\end{eqnarray}
both spinors above with positive eigenvalue, and analogously 
 \begin{eqnarray}
\label{29}\psi^{\{+,-\}}_{-}(\boldsymbol{p}) &=& \left( \begin{array}{c}
\xi_3\Lambda_{-}(p^{\mu})\Theta\phi_{L}^{\{+\}*}(\boldsymbol{0}) \\ 
i\xi_3\Lambda_{+}(p^{\mu})\Theta\phi_{R}^{\{-\}*}(\boldsymbol{0})
\end{array}  \right),\\\nonumber\\
\label{30}\psi^{\{-,+\}}_{-}(\boldsymbol{p}) &=& \left( \begin{array}{c}
\xi^{\prime}_3\Lambda_{+}(p^{\mu})\Theta\phi_{L}^{\{-\}*}(\boldsymbol{0}) \\ 
-i\xi^{\prime}_3\Lambda_{-}(p^{\mu})\Theta\phi_{R}^{\{+\}*}(\boldsymbol{0})
\end{array}  \right),
\end{eqnarray}
with negative eigenvalue. In this context we defined a complete set of eigenspinors of parity operator, the last two remaining spinors $\psi^{h}_{-}(\boldsymbol{p})$, given in equations \eqref{29} and \eqref{30}, both satisfies the dynamics presented in Section III. It is our duty to point out, that spinors in equations \eqref{27}-\eqref{30} has a non unitary transformation, in this sense, it forms a basis playing the role of expansion coefficients of a quantum field.   

\section{Dual Construction and Spin sums}
In section II, we have constructed all the possible spinors using the components given in equation \eqref{compesq} and \eqref{compdir}, the dual helicity spinors and the single helicity spinors. Now, in this present section, our next task is: use the defined single helicity spinors in Eq.\eqref{4} to find the corresponding dual (eventually check if it is the same dual as for the Dirac case as shown in \cite{ryder}). For this task we shall apply the same mathematical prescription, a mathematically rigorous recipe, as adopted recently for the Elko dual construction at \cite{1305, hep-ph}, which can be followed to construct or define any dual.  The quantity $\stackrel{\thicksim}{\psi}\;^{h}(\boldsymbol{p})\psi^{h}(\boldsymbol{p})$  must yields an invariant and real definite norm, in addition it must secure a positive definite norm for two $\psi^{h}$, and negative-definite norm for
the other remaining two. Any other choice becomes unjustifiably and forbids a physical and relevant interpretation. After some calculations, we easily conclude 
\begin{eqnarray}
\stackrel{\thicksim}{\psi}_{\pm}\;^{\lbrace +,-\rbrace}(\boldsymbol{p}) = [\psi_{\pm}^{\lbrace +,-\rbrace}(\boldsymbol{p})]^{\dag}\gamma_0,\\
\stackrel{\thicksim}{\psi}_{\pm}\;^{\lbrace -,+\rbrace}(\boldsymbol{p}) = [\psi_{\pm}^{\lbrace -,+\rbrace}(\boldsymbol{p})]^{\dag}\gamma_0.
\end{eqnarray}
The dual spinor obtained reads the same as for the Dirac case. With the dual thus defined, we have now, by construction, the orthonormality relations given by
\begin{eqnarray}
\stackrel{\thicksim}{\psi}_{+}\;^{h}(\boldsymbol{p})\psi_{+}\;^{h^{\prime}}(\boldsymbol{p}) &=& +2m\delta^{hh^{\prime}},\nonumber\\
\stackrel{\thicksim}{\psi}_{-}\;^{h}(\boldsymbol{p})\psi_{-}\;^{ h^{\prime}}(\boldsymbol{p}) &=& -2m\delta^{hh^{\prime}},\\
\stackrel{\thicksim}{\psi}_{\pm}\;^{h}(\boldsymbol{p})\psi_{\mp}\;^{h^{\prime}}(\boldsymbol{p}) &=& 0.\nonumber
\end{eqnarray}
Thus, we have defined the dual structure that provides a complete set of orthonormality relations ensuring a possible quantization.

The spin sums plays a very important role on the quantum field, it gives information about the locality structure, fermionic statistics and information about the \emph{Feynman-Dyson} propagator structure, since this quantity appears on the core of the quantum field operator. A particular aspect about the spin sums is that in some cases it is proportional to the wave operator, as it happens in the Dirac case \cite{jcap, ryder}.
So, here we will write the spin sums using the dual structure calculated, in order to obtain relevant information about the field locality structure.  
Now, we have a particular case, where the spinor dual structure matchs with the Dirac's dual.
Insomuch, is important to evaluate the spin sums, and then verify if this amount has the same structure as it has for Dirac's case. So, we have
\begin{eqnarray}\label{ss1}
\sum_{h = \{\pm,\mp\}} \psi_{+}^{h}(\boldsymbol{p})\stackrel{\thicksim}{\psi}_{+}^{h}(\boldsymbol{p}) = (\gamma_{\mu}p^{\mu} + m\mathbb{I}), 
\end{eqnarray}
and 
\begin{eqnarray}\label{ss2}
\sum_{h = \{\pm,\mp\}} \psi_{-}^{h}(\boldsymbol{p})\stackrel{\thicksim}{\psi}_{-}^{h}(\boldsymbol{p}) = (\gamma_{\mu}p^{\mu} - m\mathbb{I}).
\end{eqnarray}
Analogously with the Dirac spinors in \cite{ryder}, $\psi^{h}_{+}(\boldsymbol{p})$ can be interpreted as the bi-spinors $u(\boldsymbol{p})$ wave function and following the same reasoning for $\psi^{h}_{-}(\boldsymbol{p})$, which can be interpreted as $v(\boldsymbol{p})$ wave functions. 

\section{Final Remarks}

The purpose of this work, was to show the possibility to construct Dirac spinors, analogous as it was made for Elko, however, in abstaining parity symmetry. It should be mentioned, that in this context, we do not consider parity symmetry as a crucial link between the Lorentz representation spaces $(1/2,0)$ and $(0, 1/2)$, although this symmetry arises from the requirement to construct a complete set of spinors, as like $u(\boldsymbol{p})$ and $v(\boldsymbol{p})$, for Dirac spinor. It is important to comment about this feature, whereas we constructed the reverse path contrasting with the work \cite{speranca}. Basically, we claim that independently of choosing $P$ symmetry or the dynamics given by $\gamma_{\mu}p^{\mu}$, by means choosing one another comes naturally, and \textit{vice-versa}. Both are always correlated independent of the choice.

Guided by the classification for the spinors developed by Lounesto in \cite{lounesto}, the spinors proposed here are classified as Dirac type-$2$, when the phases are given by relations \eqref{dirac+-} and \eqref{dirac-+}, as it was expected. On the other hand, if the phases are fixed as $\xi_1 = \pm i$ and  $\xi_2 = \mp i$ we obtain spinors that does not satisfy the Dirac dynamic but belongs to type-$3$ within Lounesto classification.

We believe that parity symmetry is not an unique link between representation spaces, we have showed here that the abstention of such link does not affects the field dynamic, it remains the same as for the Dirac case. In this way, to conclude this discussion, we have showed that, if and only if, the spinor satisfy the Dirac dynamic it is an eigenspinor of parity operator, being a counterpart for what was made in \cite{speranca}.

\section{Acknowledgments}

The authors express their gratitude to professor Julio Marny Hoff da Silva for the privilege of his revision, comments and appreciation of this work. CHCV acknowledges to PEC-PG for the financial support. RJBR acknowledges to CAPES for the financial support.

\end{document}